# ASSESSING THE DIFFERENCES BETWEEN NUMERICAL METHODS, CAD EVALUATIONS AND REAL EXPERIMENTS FOR THE ASSESSEMENT OF REACH ENVELOPES OF THE HUMAN BODY

**Delangle, Mathieu; Petiot, Jean-François; Poirson, Emilie**
Ecole Centrale de Nantes, France

**Abstract**

Numerical models and computer-aided modeling software are tools commonly used to assess the accessibility of an environment, based on static human body dimensions. In this paper, the limits of validity of these approaches are assessed by comparing the reach envelopes obtained by these methods to those obtained experimentally. First, the accessibility areas of forty adult subjects were evaluated by performing a task comprising 168 reach points. Second, anthropometric characteristics of participants were recorded and used to perform the reach assessment by a numerical method, and then by a CAD-based analysis. In spite of the simple nature of the presented design problem, the results show important differences between the three methods. The study shows that the CAD-based assessment provides more accurate results than the numerical model. Nevertheless, the shapes envelopes comparison indicates that the maximum reach envelopes obtained with the CAD analysis are not always consistent with those obtained experimentally, closely linked to the hand location. Results indicate that the CAD model used to obtain maximum reaches gave predictions that underestimate the reach ability.

**Contact**:
Mathieu Delangle
Ecole Centrale de Nantes
IRCCyN
France
mathieu.delangle@irccyn.ec-nantes.fr

# 1 INTRODUCTION

The modelling of human reach envelope is widely used by designers to design and assess the accessibility of environments (for example to improve the physical design of manufacturing workplaces (Chaffin, 2007) or assess the reachability of an automated teller machine for persons in a wheelchair (Marshall et al., 2004)). Many tools and practices are based on anthropometrics data to perform these ergonomics evaluations (Roebuck, 1995). Existing database (e.g. ANSUR (Gordon et al., 1989), NHANES (NHANES, 1996) are generally chosen as the reference population. The input data are rarely measured from the real population considered for a study. To assess the accessibility of users, numerical methods, based on simple body dimensions extracted from the database, are commonly used and adapted to the environment (use case). Although numerical methods are faster and less expensive than the involvement of sample-user to test prototypes, the only use of these anthropometric data, involves a significant bias in the results of the ergonomic study (Moroney and Smith, 1972). For example, the users of a common product, independently of the age or gender, cannot always be physically compared to a military (Garneau and Parkinson, 2007). Moreover, while these data might be sufficient for univariate cases, particular care must be taken for multivariate problems, that can introduces specific complexities (Garneau and Parkinson, 2010).

Three dimensional graphical representations of anthropometric data, called digital human model, are also used to analyse and design the intended product or environment. During the past decade, software models of the human body allow an estimate of the reach envelope or the vision cone of a parameterized user in a virtual environment (Chaffin, 2001). These analyses are usually based on the manikin's inverse kinematics (IK) capability to evaluate the manikin's arm reachability in the 3D space. Human modeling tools have the ability to simulate the reach envelope of a virtual human, based on his body part dimensions and joints limits. Based on the kinematics and posture of a human, Digital Human Modelling (DHM) software are usually used instead of numerical and statistical reach models. One of the major reasons for their use seems to be that increasing the visual representation of the design problem would increase the reliability of the assessment. However, although these technologies allow a representation of the design problem in 3D environment, the ergonomic assessments are still mainly based on structural body dimensions (Brolin, 2012). Numerous DHM software include models based on static biomechanical proprieties of human body, trying to take into account factors that might affect the reach capacity as the mobility of shoulder and the spine (Chaffin, 2002). Unfortunately, these IK methods may not always be very biomechanically sophisticated and behavioural representative (Chaffin et al. 1999).

The different existing methods of reach evaluation have to allow designers to perform accessibility assessments as close as possible to the reality, without building prototypes and using participants to make experimentations. Nevertheless, it is important to know the discrepancies compared to reality that can occur using these methods.

In this study, the maximum standing reach envelopes from experiment with participants were recorded and used to assess the validity of 1) the static numerical model based on the external anthropometry, and 2) the kinematically generated reach assessments of a human model, given by the CAD modeling tool CATIA. The results obtained using a numerical method, a computer aided-design evaluation and experimental assessments are compared, and the discrepancies are discussed.

# 2 COMPARATIVE STUDY

The present study consists in comparing the evaluation of accessibility of sample-users, considering three different ways (Figure 1). The first way is to register a real experiment with a population of users. These results are used as references for the comparison. The second way consists in using the anthropometric measurements of this same population, declined with different tools:
- Numerical approach: the data are used in input of a numerical model, simulating the accessibility;
- Computer aided design approach: digital human ergonomic tools are used to predict the population's reach.

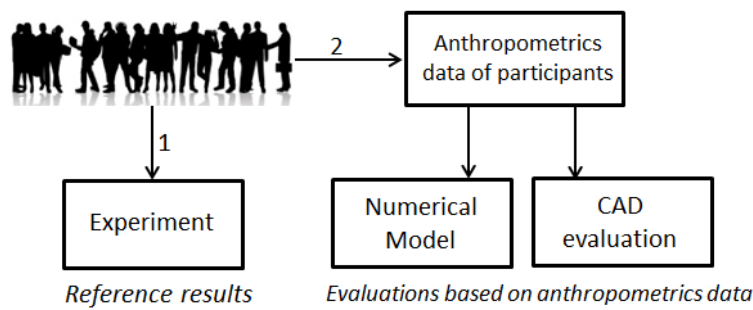

Figure 1. Synopsis of the study

*Application case*
The task proposed is a 2D reaching task of the hand of the participant in the frontal plane (Figure 2), thereof being standing in front of a wall. The aim is to compare the results found experimentally with those not involving to call a user sample given by models (given by methods commonly used in human-centred design), and thus to verify the accuracy of the different reach assessment methods.

## 3 EXPERIMENT

### 3.1 Sampling

Forty participants (twenty five males and fifteen females), adult volunteers, were sampled in the study, covering a wide spectrum of physical characteristics, from 1482 mm for the smallest stature, to 1930 mm for the highest. The average (SD) of stature of the subjects were 1735 (95.4) mm. No one of these volunteers reported motor disabilities or particular physiological limitations. The sample is considered as representative of the general population.
*It was a deliberate decision to not skew the data by "excluding" persons in the panel (e.g. old or disabled), in order to not bias the comparison. Indeed, the accessibility study of persons with specific physical limitations will depend of more parameters that would make the comparison more difficult.*

### 3.2 The test

The test is a standard accessibility situation. We seek to define the accessibility area of each user, which may correspond to the distance of reachability of products on a shelf for example. To give a target to the users, we designed a task with electric switches. A total of 84 switches could be reached on a plate, constituting 168 measurement points (Figure 2) – one high and one down point for each switch.
Participants were asked to touch the switches by colour strips (with their left or right arm according to the side), given 12 black, 32 white, 44 green, 46 blue and 34 red reachable points. The device was designed to be adjusted at the participant's shoulder height for a wide range of human physical characteristics (designed from 2.5th and 97.5th percentile for women and man stature of ANSUR database), allowing to perform the tests for a large user population. Because most anthropometric data presented in databases represent nude body measurements and to permit reliable comparison with database approach, experiments were performed with light clothing (nude dimension and light clothing being regarded as synonymous for practical purposes).

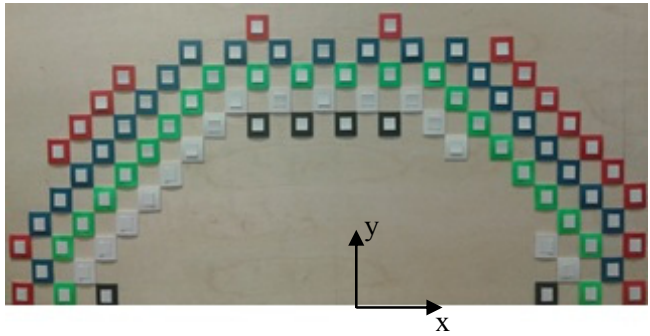
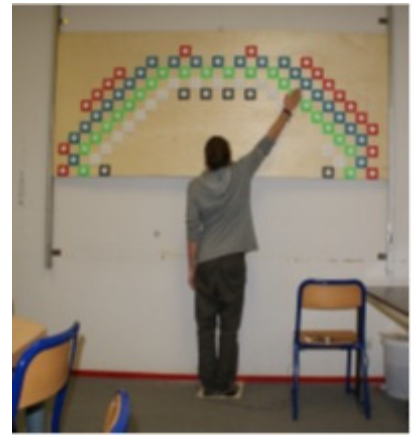

*Figure 2. Left: plate composed of 84 switches (168 points) used in the experimental trials. Right: reaching test (the subject has to reach switches on the right (left) of the plate with his right (left) arm).*

The plate was positioned on a wall with the help of two adjustable vertical axes for precise positioning of the bottom plate at shoulder height (Figure 2 Right). Reach measurements were made relatively to a body reference point (shoulder joint) and to a measurement apparatus point (bottom of the plate). The subject was positioned in the centre of the plate with the feet fixed regarding the floor. In order to check if the subjects does not lift their feet to increase their vertical reach, an electrical position sensor was positioned under their heels indicating if the feet were off the ground or not. When the visual signal was triggered, the test was stopped and the subject repositioned. This situation has to represent the functional reach, that is to say the maximal distance one can reach forward beyond arm's length while maintaining a fixed base of support in the standing position (Duncan et al., 1992), (Duncan et al., 1990).

For each individual, the reached switches were identified and noted in a table in order to draw the reach envelope.

## 4 EVALUATION METHODS

### 4.1 Creation of the common anthropometric database

When using methods for ergonomic evaluations, anthropometric modelling can be either directly observed from anthropometric characteristics of the current users, or statistically derived from characteristics for the intended target population. In order to limit the statistical biases in the comparison, the presented evaluations were all based on the anthropometric characteristics of the subjects who performed the experimental tests. External anthropometry being the type most frequently available and collected, it was decided to collect some "direct"' measurements of external link-length anthropometry. Data in Table 1 were recorded to predict the upper body accessibility for each participant, and were collected in a laboratory environment from the 40 individuals. It was expected that this number would provide a manageable database for the development and validation of the comparative study.

*Table 1. Anthropometric characteristics of the subjects (in mm) participating to the experiment, with the average and the standard deviation (excerpt).*

|                 | 1    | 2    | …   | 40   | Mean | S.D. |
|-----------------|------|------|-----|------|------|------|
| **Gender**      | M    | F    | …   | M    | -    | -    |
| **Shoulder height** | 1485 | 1450 | …   | 1420 | 1455 | 87.7 |
| **Shoulder width**  | 470  | 430  | …   | 460  | 456  | 39.8 |
| **Arm length**      | 750  | 730  | …   | 710  | 719  | 41.5 |
| **Stature**         | 1735 | 1705 | …   | 1715 | 1735 | 95.4 |

### 4.2 Static numerical evaluation

The aim is to evaluate the reach characteristics from the recorded external anthropometrics that might be found in anthropometric database (Table 1). This methodology is based on the design limits approach, which is a common method used in design problems, where data about human physical characteristics are directly applied to solve design problem. So, the maximum reach envelope is modelled (Figure 3) by an arc circle, with a radius equal to the arm length of the operator and the shoulder as point of rotation (Farley, 1955), (NASA-STD-3001, 1995). All points within this envelope (shaded area) were considered as reach by the subject. Knowing the switches coordinates on the plate and the anthropometrics characteristics of individuals, a program was implemented allowing to automatically determinate which switches the participants theoretically reached. The results are presented in section 5.

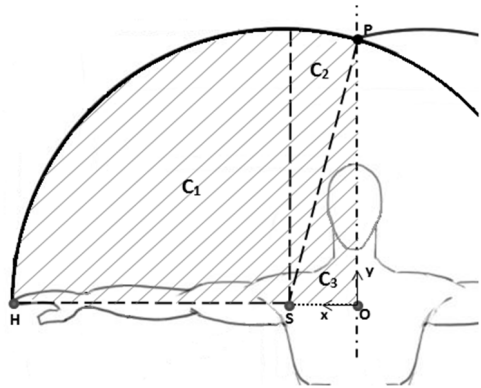

*Figure 3. Static model of reach envelope (shaded area) of the arm calculated from external anthropometric dimensions of the subject.*

### *4.3 Computer aided evaluation*

The use of computer aided ergonomics systems during product development is a well-established methodology of which there are numerous examples (Marshall et al., 2004), (Case et al., 2001), (Das and Sengupta, 1995). Many software, all providing different manikins model and ergonomics tools, are available to the designer (Delangle and Poirson, 2013), (Ranger, 2010). A Human simulation technology commonly provided is the assessment of the reach envelope of a digital human model (DHM) within a digital mock-up (DMU) (Chaffin, 2002). In numerous situations, and particularly for multidimensional design problems, these simulations, based on the kinematics and posture of a human figure model are used instead of statistical reach models. Nowadays, a lot of fields of engineering, as in automotive (Hanson and Högberg, 2008), (Yang and Abdel-Malek, 2009) or in the inclusive design area (Marshall et al., 2010), (Porter et al., 2004), use these technologies to predict the reachability of an environment.

This section offers to assess the validity of the kinematics reach envelope generated with a digital human model. CatiaV5 and its Human modeling module Delmia were used to assess the reachability of the environment. This software uses the manikin's inverse kinematics capability to evaluate the manikin's arm reachability in the 3D space (joints ranges of motion of the manikin). The maximal reach envelope, including spine bending, is created to represent the maximum reach limits. As for the previous method, the evaluation was performed based on the experimental recorded data (Table 1). The virtual manikin was parameterized on the model of each participant. The model of user trials that was performed with real people was taken and moved into the virtual world. The position of the avatars and the dimensions of the problem have been specified by the designer to be completely representative of the real situation. A display of the reach assessment is depicted in Figure 4. Results are discussed section 5.

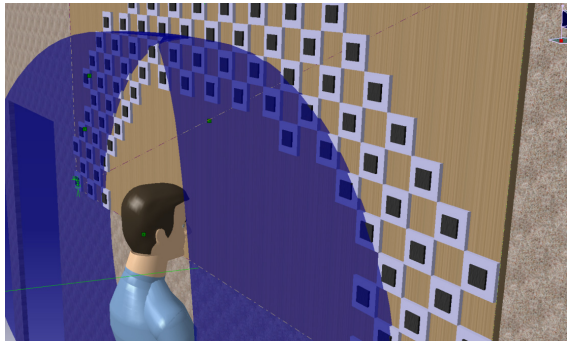

*Figure 4. CAD evaluation of the maximal reach envelope of a participant.*

# 5 RESULTS / DISCUSSION

## 5.1 Global comparison

Results from experiments and evaluations were collected for each of the forty participants. To assess the global reach, the number of switches reached by each participant was recorded. For each method, the mean and the associated standard deviation were calculated for each evaluation (Table 2).

*Table 2. Numbers of reached switches for each participant, measured from the experiment and predicted through numerical and CAD method, with means and standard deviations (excerpt).*

|  | 1 | 2 | … | 40 | Mean | S.D. |
|---|---|---|---|---|---|---|
| **Experiment** | 107 | 102 | … | 52 | 96 | 25 |
| **Numerical model** | 76 | 57 | … | 30 | 63 | 22 |
| **CAD evaluation** | 87 | 67 | … | 44 | 79 | 26 |

Intrinsically to the experiment, the overall reach envelope is highly correlated with the stature of the subject. The higher the stature, the higher the number of reached switches. In order to highlight the differences between the evaluations (without the stature influence), results were normalized by the stature of the participant. Resulting means and standard deviations obtained are depicted in Figure 5. This graph allows the comparison of the global results of each evaluation.
The static numerical model appears to be the method that underestimates the most the results, with an average of 36 reached points. Then, the assessment performed with the computer aided ergonomics tool provides more accurate results compared to the experiment, with an average of 45 switches. Comparing with that obtained experimentally (55 switches), the static numerical and CAD evaluations represent respectively 64% and 82% of the total reach really obtained. The lowest standard deviation (Table 2) corresponding to the static numerical evaluation, represents the disparity of reach only related to differences in anthropometric characteristics.
In spite of the simple nature of the presented task (two-dimensional), the results show important differences between the two evaluations. The reach capacity is not only correlated with the static dimensions but depends on other external variables. This comparison being done for the overall reach, it is involved that a subject could reach more switches as expected (regarding his physical characteristics) on certain parts of the envelope, and less than expected in others. That's why, it addition to the global results, the reach was also studied in a local way.

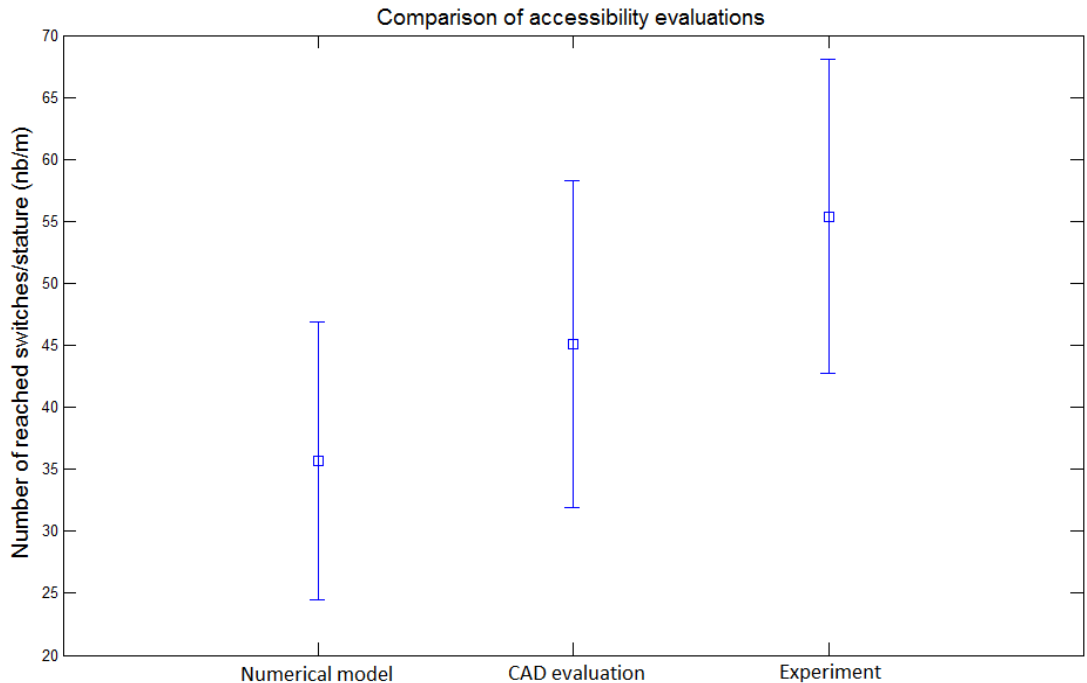

*Figure 5. Mean and standard deviation of the number of reached switches divided by the stature for the 40-member sample, obtained from the four methods*

## 5.2 Local comparison

The CAD assessment provides results the closest to the experiment, with a total error of 672 against 1330 for the numerical evaluation (Table 3). The total errors are defined as the differences between the total numbers of points reach using each evaluation methods and the experiment, representing the deviations from the "reference" result (considered arbitrarily as the results of the experiment).

To study the localizations and frequencies of these errors, a local comparison was conducted. The normalized theoretical reached envelope obtained from static numerical evaluation was used as a basis for comparisons to highlight the behaviour of the reach depending on the hand position during the movement. To avoid the anthropometric differences, each envelope was normalized between 0 (corresponding to the initial shoulder position) and 1 (dimensional extremity of the hand). The reach errors were determined for each participant and aggregated for each method. For example, shown in Figure 6 is the aggregate error for each zone of the envelope, obtained from the CAD evaluation. The grid corresponds to the discretization of the domain, representing the different points (switches) reachable on the plate. A positive error (on the vertical axis) corresponds to unexpected reached points (outside the theoretical arc circle), and negative error to unreached points that were expected to be reached (into the theoretical arc circle).

In order to make the comparison between the evaluations, the reaches were drawn on a two dimensional normalized graph (Figure 7). The reach envelopes defined from the static numerical evaluation are represented by an arc circle, with 0 as centre and 1 as radius. The shaded area represents the theoretical reach envelope based on the numerical model. Any point include into this arc circle is theoretically reached. Conversely, any point outside this area is theoretically considered as unreached, considering the body structural dimensions. The lines represent the mean profile of the reaches for the forty subjects, i.e. (50% included and 50% excluded). Thus, all the parts of a line not included in this area describe the set of points reached out of the theoretical reach.

The maximum vertical reach (corresponding to points above the head) is higher than expected for the experimental evaluation. The main reason of this increase is probably due to the complex model of displacement of the shoulder, and particularly of its flexion, allowing the subjects to increase their reach upwardly, beyond their external structural dimensions. With the foot fixed on the ground, the strategy to enhance the vertical reach is to increase the flexion of the shoulder and to change the alignment of the shoulder and the pelvis. The same observation can be made for the CAD evaluation, where the maximal extended reach is located beyond the head. This might be explained by the 5 degrees of freedom shoulder model of the CATIA mannequin, allowing elevation / depression movement, i.e. an upwardly augmentation of the reach (Ranger, 2010), (Dassault-Systemes, 2000).

Table 3. Total number of reached switches for each participant (excerpt) and the reach error compared to the experiment results.

|  | 1 | 2 | … | 40 | Total | Error |
|---|---|---|---|---|---|---|
| **Experiment** | 107 | 102 | … | 52 | 3846 | 0 |
| **Numerical model** | 76 | 57 | … | 30 | 2516 | 1330 |
| **CAD evaluation** | 87 | 67 | … | 44 | 3174 | 672 |

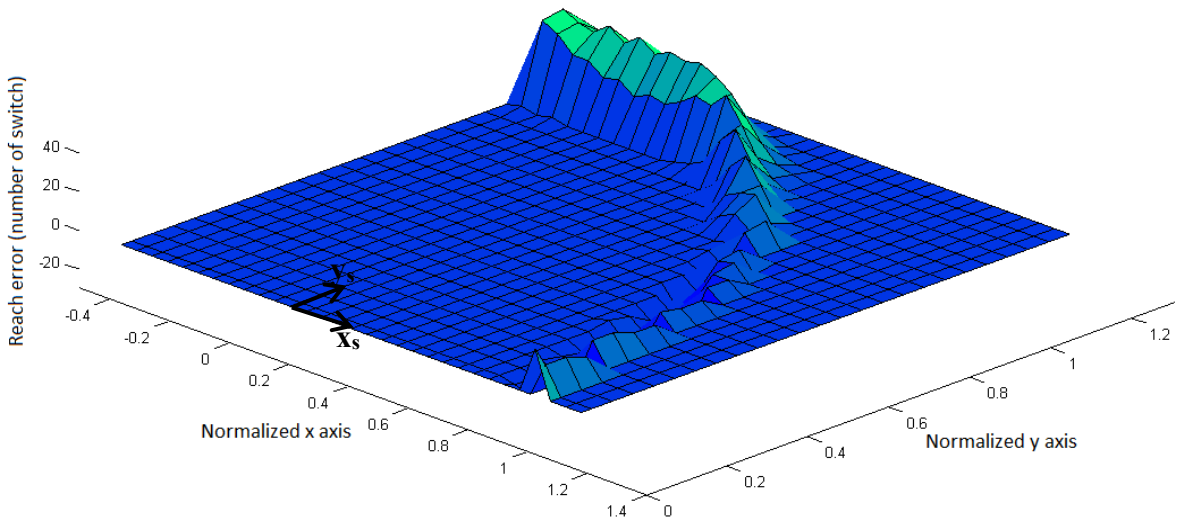

Figure 6. Plot of the overall model error in the various regions of the reach envelope, obtained from the CAD evaluation (with the numerical model as reference). The frame represents the initial right shoulder position in the normalized domain.

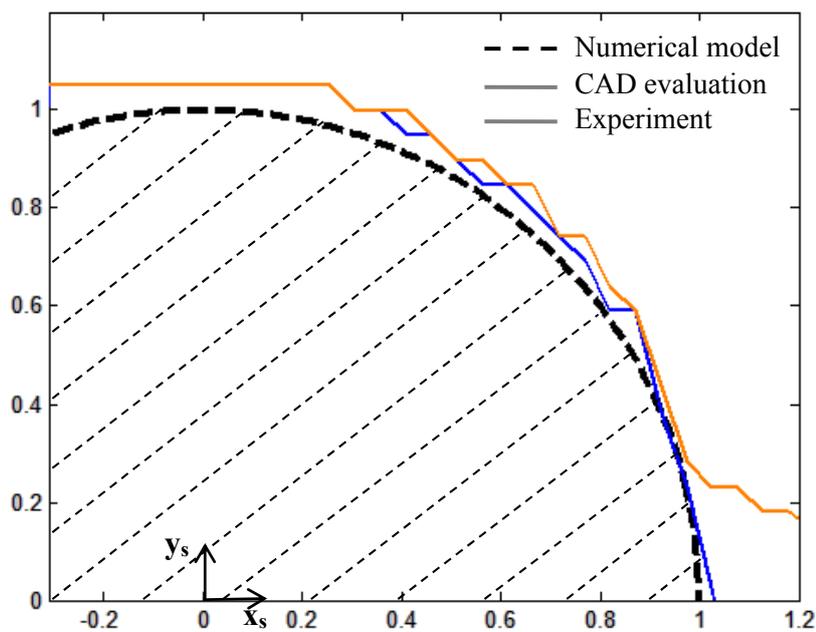

Figure 7. Normalized reach lines (50% accommodation) obtained from the experiment (orange) and from the CAD evaluation (blue). The shaded area represents the normalized reached area obtained from the static numerical model. The frame represents the initial right shoulder position in the normalized domain.

According to Figure 7, the reach envelopes obtained experimentally appear to be overall more important than those predicted by the models. Nevertheless, the greatest difference is localized on the lateral region of the envelope. The body being not constrained (excepted the feet), the maximal reach strongly depends on the lateral reach capacity, represented by the medio-lateral balance stability and the pelvis rotation ability of the subject (Brauer and Burns, 1999). It appears that it is quite difficult to take into account this kind of behaviour using only a computer-aided design environment. Most users of digital human models will attempt to overcome these limitations by kinematic manipulations of the human model. However, due to a lack of quantitatively valid models of postures and motions for extreme reaches, the results may not be useful. For example, a user of a digital human model is unlikely to guess the correct amount of lateral pelvis rotation when simulating lateral reach aspects (Abdel-Malek and Arora, 2009), (Reed et al., 2003).

The CAD assessment, including biomechanical considerations intrinsic to the human model of the software, made more realistic results compared to a model only based on the static anthropometric dimensions. Nevertheless, this model seems to still underestimate the range of extended reach involved by the human behaviour. This is consistent with the general problems associated with Human-CAD models of reach, which might have difficulty to predict reaches in the outer borders of the reach envelope (Kozey, 2002).

## 6 CONCLUSION

This study compared results of a reach assessment obtained from an experiment, a numerical model and a computer aided evaluation. In spite of the simple nature of the presented design problem (two-dimensional), the results show important differences between the three evaluations. Those obtained using the CAD evaluation were the most accurate, compared to those really obtained from the experiment. However, the local comparison showed that the reach envelopes provided by the CAD assessment were not always consistent with those obtained experimentally, depending on the location of the hand. The results indicate that the CAD model used to obtain maximum reaches gave predictions that are likely to be substantially in error. The majority of the discrepancy between the observed maximum reach capability and that predicted by the model is located on the lateral reach area, and is particularly associated with the medio-lateral balance stability and the pelvis rotation ability of the subject. This means that a strictly kinematical approach to predicting maximum reach capability is not likely to be accurate.

Anthropometric design problems associated with human physical characteristics depend on many other factors. Several task considerations should be taken into account in order to construct a reach envelope, as the nature and requirements of the task to be performed, the body position while reaching, the whole body movement capabilities and restraints. Thus, human models included in the DHM'S cannot include all the subtleties of human movement observed in the large variety of coping strategies that may be used to achieve success. This is particularly true for more complex evaluations, where the actual posture adopted may be important, such as restricted access tasks.

Human models used for reach assessment should consider all these human capacities. A solution could be to numerically include biomechanical and behavioral data in the inverse kinematics model, in order to take into account the extended reaches. Unfortunately, this will be difficult to apply in the case of complex multivariate design problem, due to the wide variety of human capacities and reach strategies involved. That is why design evaluation should be made in order to naturally take into account these strategies. Thus, virtual reality tools, allowing a designer to be physically involved in the simulation, seem to be a promising approach in the field of reach evaluation in design. Indeed, because these new technologies allow the reproduction and simulation of real movement of a human interacting with an environment, virtual simulations might take into account more factors influencing the reach that existing method. Development of virtual reality technologies could be a key approach to improve the user-centred design methodology, especially for complex multivariate problems, to develop a powerful ergonomic design tool, including biomechanical, structural, behavioural and cognitive models.